# A Meta-Analysis of Solar Forecasting Based on Skill Score

Nguyen, T.N.[a,1,*], Müsgens, F.[a]

[a] *Brandenburgische Technische Universität Cottbus-Senftenberg, Siemens-Halske-Ring 13, 03046 Cottbus, Germany*

**Abstract**

We conduct the first comprehensive meta-analysis of deterministic solar forecasting based on skill score, screening 1,447 papers from Google Scholar and reviewing the full texts of 320 papers for data extraction. A database of 4,687 points was built and analyzed with multivariate adaptive regression spline modelling, partial dependence plots, and linear regression. The marginal impacts on skill score of ten factors were quantified. The analysis shows the non-linearity and complex interaction between variables in the database. Forecast horizon has a central impact and dominates other factors' impacts. Therefore, the analysis of solar forecasts should be done separately for each horizon. Climate zone variables have statistically significant correlation with skill score. Regarding inputs, historical data and spatial temporal information are highly helpful. For intra-day, sky and satellite images show the most importance. For day-ahead, numerical weather predictions and locally measured meteorological data are very efficient. All forecast models were compared. Ensemble–hybrid models achieve the most accurate forecasts for all horizons. Hybrid models show superiority for intra-hour while image-based

[1] *Corresponding author

*Email address:* nguyen@b-tu.de (Thi Ngoc Nguyen)

Correspondence to: Brandenburgische Technische Universität Cottbus-Senftenberg, Siemens-Halske-Ring 13, 03046 Cottbus, Germany.

methods are the most efficient for intra-day forecasts. More training data can enhance skill score. However, "over-fitting" is observed when there is too much training data (> 2000 days). There has been a substantial improvement in solar forecast accuracy, especially in recent years. More improvement is observed for intra-hour and intra-day than day-ahead forecasts. By controlling for the key differences between forecasts, including location variables, our findings can be applied globally.

---

**Abbreviations**

| | |
|---|---|
| ANN | artificial neural network |
| CP | a convex combination of smart persistence and climatology models |
| CZ | the first level classification of climate zone based on Köppen-Geiger. |
| CZA | equatorial climate zone |
| CZB | arid climate zone |
| CZC | warm temperature climate zone |
| CZD | snow climate zone |
| CZE | polar climate zone |
| CZN | no information of climate zone |
| DL | deep learning |
| DNI | direct normal irradiance |
| Ens | ensemble models |
| Ens_Hyb | ensemble–hybrid methods |
| GHI | global horizontal irradiance |
| ICS | clear sky irradiance |
| InputHist | using the historical values of solar power as among the input variables |
| InputMete | using the locally measured data of meteorological variables as among inputs |
| InputNWP | using the numerical weather predictions as among inputs |
| InputST | using spatial-temporal variables (e.g., from neighbor power sites) as input |
| KG | the Köppen-Geiger classification of climate zones |
| MAD | median absolute deviation |
| MAE | mean absolute error |
| MARS | multivariate adaptive regression spline |
| ML | machine learning |
| ModClass | the classification of models |
| MSE | mean squared error |
| N | number of observations |
| NWP | numerical weather predictions |
| PDP | partial dependence plot(s) |
| pp | percentage point |
| PV | photovoltaics |
| ResMin | the forecast resolution [minutes] |
| RMSE | root mean square error |
| SD | standard deviation |
| SE | standard error |
| SP | smart persistence model |
| SS | skill score [%] |
| SVM | support vector machine |
| TestLength | the length of the test or validation data [days] |
| TrainLength | the length of the training data [days] |
| Trim | trimmed mean value |
| TS | time series method |
| TWh | terawatt hour |

## 1. Introduction

In the last decade, global solar energy generation has increased exponentially from 63.8 TWh in 2011 to 821 TWh in 2020 [1]. It ranked second among renewable technologies in terms of absolute generation growth in 2020, slightly behind wind and ahead of hydropower. Solar power remains the lowest-cost option for electricity generation and contributes substantially to the goal of net zero emissions. The International Energy Agency's tracking report on solar power [1] indicates that more efforts are needed to achieve 7,000 TWh of solar energy by 2030.

Such rapid growth in global solar energy generation may increase uncertainty in the feed-in of renewable energy [2]. Dealing with this uncertainty is crucial for reliable power system operation [1] and solar (power) forecasting emerges as a particularly efficient solution [3]. More accurate forecasts enable the system to plan power generation more efficiently and match it with consumption at lower operational cost. This also allows the integration of a higher share of solar energy into the system. Furthermore, desired levels of reliability in the grid can be achieved at lower cost, for example by balancing power procurement. Therefore, there is a high potential business value in accurate forecasting.

As a result of these potential benefits, academic researchers have published hundreds of papers on enhancing the accuracy of solar forecasts. The accumulated volume of research in the field and rapid expansion of methodologies [4] leads to a need for systematizing the scientific knowledge. However, the task is complicated by the diversity of the datasets and forecast setups [5]. Many studies show that factors such as weather conditions, forecast horizon, and the quality of the input data can significantly influence forecast accuracy [3]. These factors should be accounted for when conducting inter-model comparisons. In addition, variability in forecast validation approaches also causes difficulties in comparing results [6].

The best method to address such differences when comparing forecasts is through meta-analysis. A meta-analysis is a statistical approach to extract findings from individual studies using quantitative methods [7]. For example, a regression can be conducted on the forecast accuracy data and the effects of all relevant variables controlled. From the output of this regression, different forecast methods can be compared and the impacts of different factors are analyzed. Such insights are important for both industry and academia to adopt the best practices in forecasting. Furthermore, by accounting for the impacts of location contingent variables (e.g., climate conditions), the analysis results can be applied globally. This paper provides the first comprehensive meta-analysis of deterministic[2] solar forecasting using the skill score (SS) metric.

SS is recommended as a standard metric to measure forecasts' performance [8]. Due to its relative measurement approach, this metric reduces the impact of inherent difficulties in different forecasting situations and renders forecasts across studies more easily comparable [9]. While this property of SS makes a meta-analysis of solar forecasting more convenient, an in-depth analysis of forecasts and an extensive inter-model comparison based on the metric is almost unknown. The volume of studies reporting this metric has increased so substantially that a meta-analysis based on SS is now feasible.

The work in this paper contains novel contributions both in methodology and empirical findings:

- Methodologically, the paper provides the first complete meta-analysis of deterministic solar forecasting based on the SS metric. We screen 1,447 papers from Google Scholar, among which the full texts of 320 papers from 2006 to 2022 are thoroughly reviewed for data extraction. A database of 4,687 observations for 11 variables is built, and the non-linear relationships and interaction terms between the variables analyzed through multivariate adaptive regression spline modelling (MARS) and partial dependence plots (PDP). Key MARS results are then used to derive additional insights from linear regression.

[2] We focus on deterministic forecasts, also referred to as point forecasts, due to their greater applicability and the volume of the literature.

- Empirically, the analysis provides valuable insights that can be applied globally to improve forecast quality. Several variables are added whose impacts on forecast accuracy are rarely discussed and quantified in the literature. In addition to variables such as forecast horizon and forecast models, other variables including forecast resolution, climate conditions, train and test sets, and the use of different inputs are also considered. An overview of scientific progress in the field is also provided.

To the best of our knowledge, only two meta-analysis studies have been conducted on solar forecasting. In both studies, however, the analysis was made on normalized errors rather than SS. The first by Blaga et al. (2019) [10] focused exclusively on solar resource forecasting and the second by Nguyen & Müsgens (2022) [5] covered photovoltaic (PV) output forecasting. Blaga et al. [10] extract data from 40 papers between 2007 and 2016 and analyze inter-model performance. They consider the effects of different factors such as forecast horizons and climate conditions. Using data visualization, the work generalizes findings from individual studies and provides important insights into solar resource forecasting. In Nguyen & Müsgens (2022) [5], the data is extracted from 69 papers on PV output forecasting. In addition to data visualization methods, a regression is also conducted to analyze the impacts of variables on the forecast errors of models. The impacts of forecast horizon, test set length, techniques, and inter-model comparison are analyzed. In our work, a meta-analysis of solar forecasting based on the SS metric is provided. Furthermore, we extended both the number of observations and number of variables in the database. Our work is based on 4,687 data points extracted from 188 papers on solar irradiance or resources and PV output forecasting. Each observation is represented by the SS value and 10 other dimensions of features, which represent the key differences between forecasts.

The structure of the paper is as follows. Section 2 describes the data extraction and meta-analysis methods. Section 3 presents the results and discussion. Section 4 concludes the paper.

## 2. Methods

This section explains the background ideas for building the database of solar forecast accuracy and the meta-analysis methods.

### *2.1. Background for data extraction*

The key objective of solar power forecasting is to provide accurate predictions of "solar power" before its realizations can be observed. Hence, we start with a discussion of forecast accuracy and verification in Section 2.1.1. We also discuss the SS metric as the standard indicator to measure forecast accuracy, i.e., the dependent variable in our analysis. In Section 2.1.2, central factors that can drive the accuracy of forecasts are discussed, i.e., the independent variables in our analysis.

#### *2.1.1. Forecast accuracy measurement indicators*

Currently, the accuracy or quality of forecasts is measured using a variety of different indicators or metrics. D. Yang et al. (2018) [11] show that there are at least 18 metrics to validate deterministic forecasts. Among these, root mean square error (RMSE) is the most widely used. The RMSE metric is more sensitive to spikes in data (e.g., severe solar ramps) due to the squared values. This is desirable if model accuracy in extreme events is a requirement. Among the other metrics, mean absolute error and mean bias error are also frequently used [10].

The rapid expansion of methodologies for solar forecasting [4] leads to the need to compare forecasts across datasets. This requires scale-independent metrics [11], which can be achieved through normalizing with power generation in the denominator [12]. Typically, power generation can be the average or weighted average power, the peak nominal irradiance or peak PV power, or installed capacity [13]. Note that changing the denominator can lead to large changes in the values of the normalized errors [5]. Therefore, the denominator options should be accounted for when comparing the normalized metrics across studies.

Additional factors such as geographical and weather conditions should also be considered when comparing forecasts [14]. The forecast skill or skill score (SS) metric effectively addresses this [8]. SS measures the performance of a model by comparing its accuracy with a specific reference model. This relative measurement [15] takes into account the variability and uncertainty of the forecast situation and allows models to be comparable. The property makes SS a natural candidate for a meta-analysis of solar forecasting, where an inter-model comparison is desired. Therefore, the analysis in this paper focuses exclusively on SS. The SS values reported in the literature are extracted and represent the dependent variable in the database.

SS can be calculated as follow:

$$SS = \frac{A_f - A_r}{A_p - A_r}, \quad (1)$$

where $A_f$, $A_p$, and $A_r$ are the accuracy measurements (e.g., mean squared errors) of the forecasts of interest, the perfect forecasts, and the reference forecasts, respectively [16]. In solar forecasting, it is often assumed that the accuracy measure of a perfect forecast $A_p$ is much smaller than the reference forecast's $A_r$ and can be approximately 0. Therefore, SS can be reformulated as:

$SS = 1 - A_f / A_r$. A positive value of SS indicates that the forecasts of interest have higher accuracy than the reference forecasts, i.e., the higher SS, the better.

To calculate SS, scholars need to decide on the accuracy measurement metric and the reference method. There are different suggestions for the accuracy measurement metrics. For example, Perez et al. (2013) [17] used mean squared error (MSE), Morf (2021) [18] recommended using Spearman's ρ as the skill score, etc.

This paper follows the majority of the literature which calculates SS based on the RMSE metric. As data for the reference method are more disperse, we include the three most widely used methods: skill scores based on persistence[3] ($SS^P$), based on smart persistence[4] ($SS^{SP}$), and based on a convex combination of smart persistence and climatology[5] models ($SS^{CP}$). $A_r^x$ in the respective $SS^x$ is calculated as follows, given that the accuracy measurement is RMSE:

$$A_r^P = \sqrt{\sum_{i=1}^{n} (x_i - y_i^P)^2 / n}, \text{ with } y_i^P = x_{i-1}, \forall\, i = 1,2,\dots,n \text{ and assuming } x_0 \text{ is defined.} \quad (2)$$

In $A_r^{SP}$, $y_i^P$ is replaced by $y_i^{SP} = y_{i-h} * \frac{ICS_i}{ICS_{i-h}} = k_{i-h} * ICS_i$, with $ICS$ being the clear sky irradiance and $h$ denoting the values h-steps prior.

[3] The persistence model assumes that the value for solar power persists over a certain period and issues the forecast as the most recent observation.
[4] Smart persistence is similar to persistence but uses the clearness index (k) to correct the forecasts.
[5] The climatology model generates a constant value for all forecasts. The constant value is calculated as the average value of the whole sample.

Lastly, for $A_r^{CP}$, $y_i^{CP} = \alpha y_i^{SP} + (1-\alpha) y_i^{climatology}$, where $0 < \alpha < 1$ and $y_i^{climatology} = \bar{x}$ for all time steps $i$, with $\bar{x} = (1/n)\sum_{i=1}^{n} x_i$, and $i = 1,2,\ldots,n$.

To analyze the database of $SS^P$, $SS^{SP}$, and $SS^{CP}$, the data visualization is plotted separately for each reference model. In the regressions, the variable of the reference model is included so that the effects can be accounted for, and the three SS values can be analyzed collectively.

#### *2.1.2. Factors influencing forecast accuracy*

Through our review of historical studies and literature surveys on solar forecasting (see Table A.1), we show that the top ten important variables that significantly influence forecast accuracy are (alphabetically) climate conditions, forecast horizon, inputs, models of forecasts, reference model (used to calculate skill score), resolution of forecasts, test set length, train set length, the type of forecasts (solar resource or PV output forecasts), and the year of publication of papers (more details on this in Appendix A). These variables' impacts on forecast accuracy should be considered when analyzing or comparing forecasts' performance. To that end, the information of these variables is extracted from the reviewed papers to build a database of solar forecast accuracy, with the dependent variable being SS, and the independent variables being these factors. Details of how the ten independent variables are extracted are discussed in Appendix A. The database is summarized in Table 1. In the next section, the meta-analysis process is explained in more detail.

Table 1: Description of the variables in the database. The variables are ordered alphabetically.

| No | Variable | Description |
|---|---|---|
| 1 | CZ* | The first level classification of climate zone based on Köppen-Geiger. |
| 2 | Horizon | The length of the forecast horizon [minutes] |
| 3 | InputHist** | Using the historical values of solar power as among the input variables |
| 4 | InputMete** | Using the locally measured data of meteorological variables as among inputs |
| 5 | InputNWP** | Using the numerical weather predictions as among inputs |
| 6 | InputST** | Using spatial-temporal variables (e.g., from neighbor power sites) as input |
| 7 | ModClass* | The classification of models |
| 8 | Reference* | The reference model used to calculate SS |
| 9 | ResMin | The forecast resolution [minutes] |
| 10 | SS | The value of skill score reported for the model of interest [%] |
| 11 | TestLength | The length of the test or validation data [days] |
| 12 | TrainLength | The length of the training data [days] |
| 13 | Type* | The type of the forecasts, i.e., PV output or solar resource forecasting |
| 14 | Year | The publishing year of the paper |
| *Note: *: categorical variable, **: dummy variable, otherwise: numerical variable* | | |

### *2.2. Meta-analysis method*

The meta-analysis was conducted in five steps as illustrated in Figure 1. The first three steps, from identification to full text review, describe the literature selection process and are presented in Section 2.2.1. The fourth step presents the extraction and cleaning of the database (Section 2.2.2) and the final step presents the data analysis (Section 2.2.3)

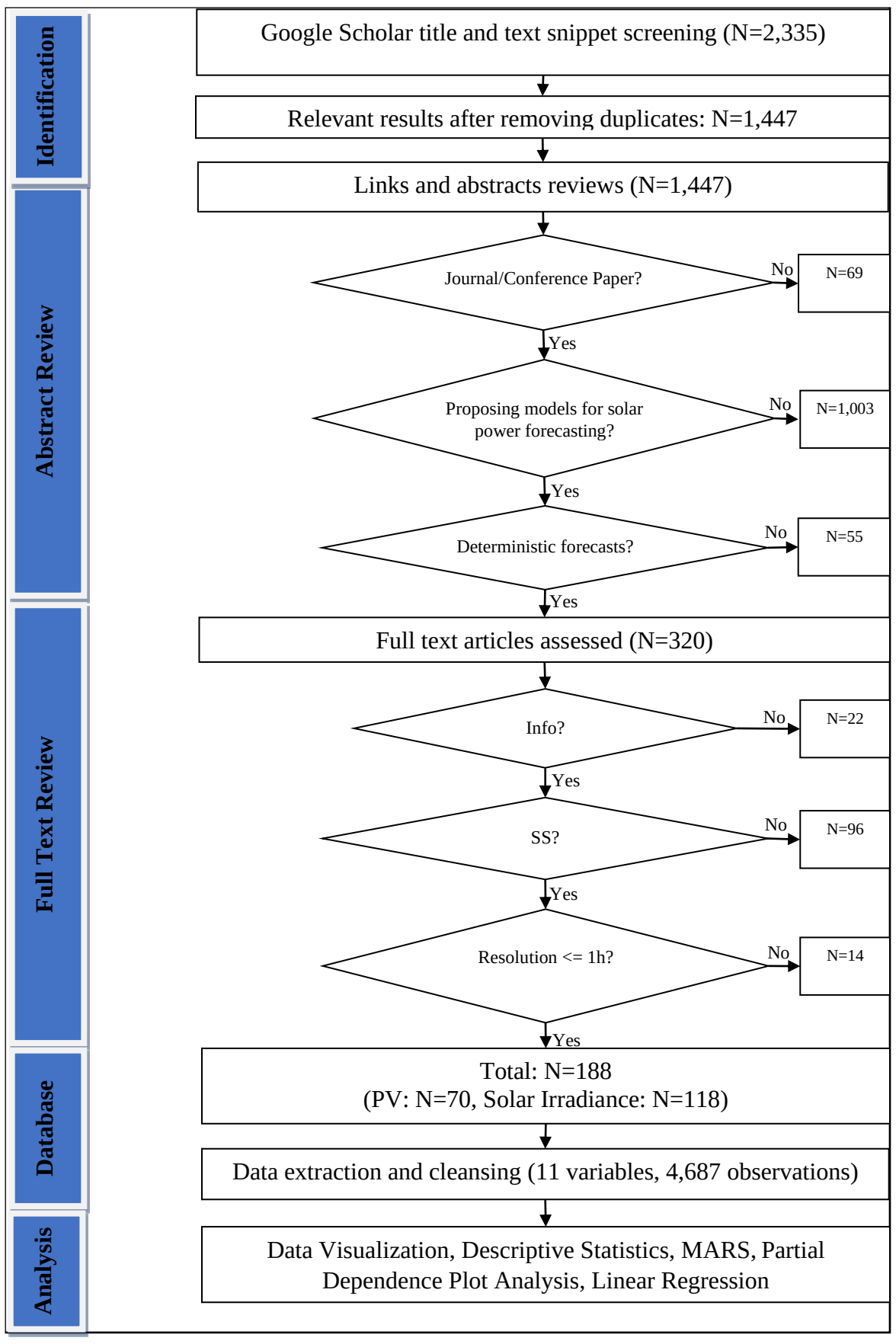

Figure 1: The meta-analysis process illustrated in five steps.

### *2.2.1. Literature selection*

The first step in the literature selection process was literature identification based on a comprehensive search of Google Scholar. We performed four search sessions using different combinations of keywords, which are summarized in Table B.1. A total of 2,335 results were screened based on the title and text snippets (the short text description under the title). If the title or the text contained solar resource or PV forecasting equivalent terms, and the term “skill score” (or “forecast skill”) was mentioned, the result was considered relevant. In cases of doubt, the result was also included as relevant at this stage. At the end of this step, 1,447 unique search results were considered relevant and moved to the next step. The list of 1,447 articles is provided in the supplementary data file.

The second step was a review of the abstracts in which the 1,447 relevant results were checked for more details. Three criteria were checked:

i) Is it a journal or conference paper?

ii) Is it proposing a model for solar power forecasting?

iii) Is it about deterministic forecasts?

When all three questions were answered in the positive, we kept the articles for further processing. Question (i) removed 69 results including materials such as books, theses, and reports. An additional 1,003 results did not satisfy question (ii). Among them, some were review papers, others were about different fields such as weather forecasting, wind forecasting, power system optimizing, etc. Finally, 55 results were excluded because they provided probabilistic rather than deterministic forecasts, question (iii). Overall, out of 1,447 results, 320 papers remained for full text assessment.

The third step was a full text review of the eligibility of the data. Only articles in which all three of the following questions were answered in the positive were kept:

i) Is the information adequate and clear?

ii) Is an SS based on RMSE and a reference method of persistence, or SP, or CP reported?

iii) Is the resolution for each forecasted time step less than or equal to 1 hour?

Regarding question (i), some papers did not provide sufficiently clear information. For example, some papers did not specify whether daily or hourly forecasts were performed, others did not specify forecast horizons. Many papers were unclear about data partition to perform out-of-sample evaluation. The following key information was required from the papers: the forecast horizon and forecast resolution, the test set, and a clear explanation of the SS calculation including choice of metric and reference method. There were 22 papers excluded for not providing the required information.

Regarding question (ii), several papers calculated the SS based on other metrics, such as MAE or MSE. Furthermore, a variety in the choice of reference methods was observed. While the majority of studies used one of the three reference methods discussed in Section 2.1.1, it does not hold for all. As discussed, the meta-analysis in this paper focuses on the SS calculated by RMSE and reference methods of persistence, SP, and CP. All the other formulations of SS were not included. With this criterion, 96 more papers were excluded.

Regarding question (iii), the resolution of the forecasts was restrained. Researchers can provide forecasts for the PV output in resolutions ranging from every second to every day, or even to every month. Among the studies reviewed, resolutions tended to be high (forecasting the PV output every hour, half-hour, or shorter, for example). Some studies forecast at a lower resolution, in particular the average power per day or half-day. This is less complicated than hourly forecasts and accounts for an insignificant proportion of observations. Fourteen papers were excluded.

The remaining 188 papers were kept for data extraction. The list of these papers is presented in Appendix C.

### *2.2.2. Database*

Step four sets up the database from the 188 papers. Note that one publication often provides more than one observation as different models are presented and forecasts can be for different locations, horizons and resolutions. In total, we derived 4,687 observations from the papers. Each observation contained 11 variables, including SS as the dependent variable.

After the data extraction, additional steps were conducted to fix data formats. First, all differences in units were harmonized. The data type for each variable was then set according to their values and all numerical values formatted consistently. Finally, the whole database was checked for missing values and corrected for other formatting mistakes. A summary description of the variables is presented in Table 1. For numerical and dummy variables, important statistics are summarized in Table D.1.

The allocation of data over climate zones is illustrated in Figure 2. More than half the database belongs to climate zone C. Climate zone B covers around 20% of the data. The remaining data (around 22%) comprises the other climate zones. It might be wondered whether what works best in the climate zone C also works best in other areas. The possibility of transferring insights between regions to improve forecast accuracy is crucially important, especially for those whose research on solar forecasting has not been advanced. By accounting for the impacts of different factors simultaneously, the meta-analysis in this paper enables this transfer of knowledge.

More details about the database are presented in Appendix D.

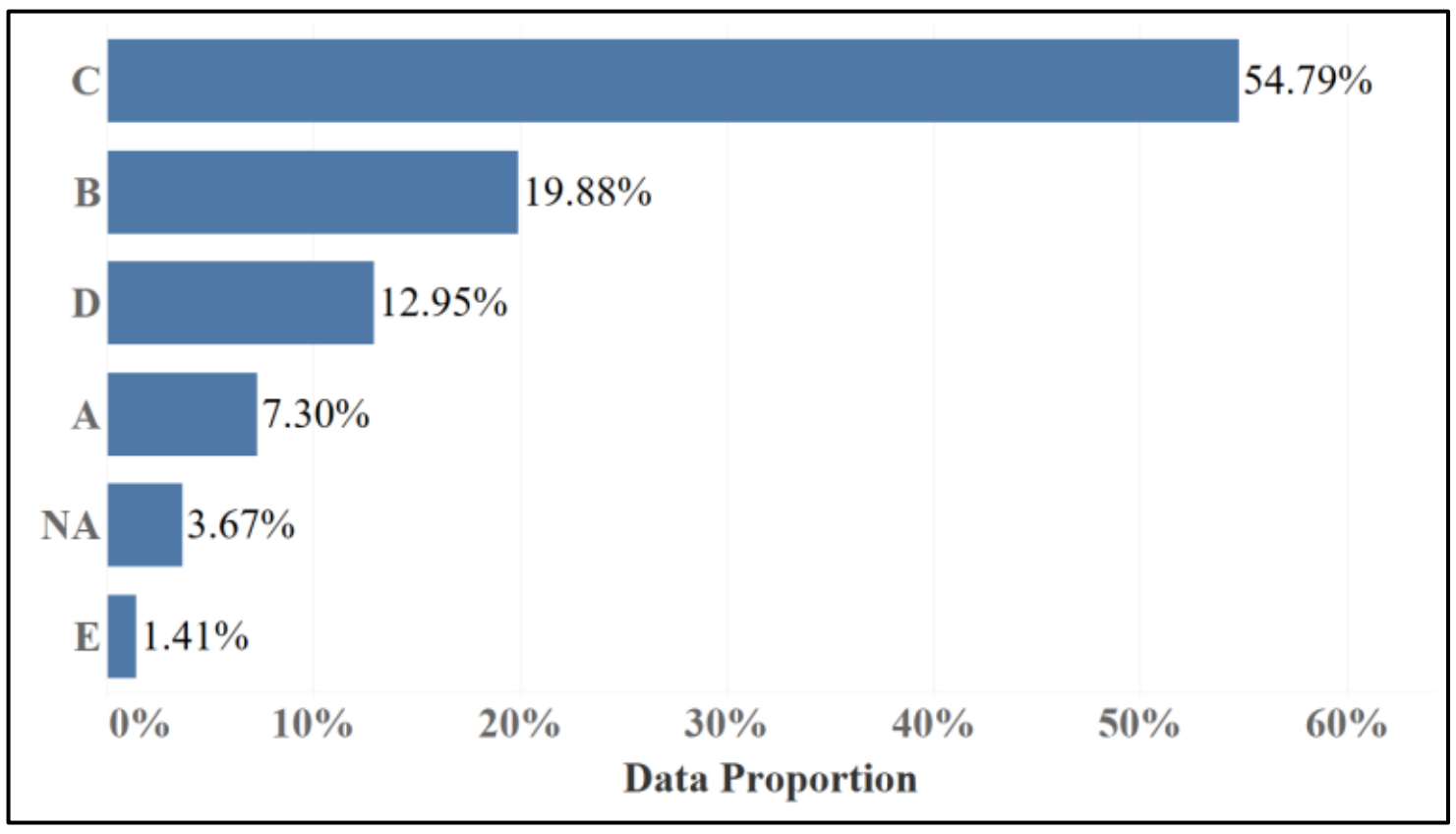

Figure 2: Data distribution by climate zone.

#### 2.2.3. Model

A meta-analysis requires a methodology to measure correlations and compute quantitative results. Regression methods were applied because they allow examining the impact of one variable while controlling for the impacts of the other variables. In the first place, multivariate adaptive regression spline modelling (MARS) and partial dependence plots (PDP) were applied to examine the interactions of variables in the database. Through these methods, the most important interactions were identified and analyzed, allowing for non-linear relationships in the process. Based on this, the database was then split into partitions based on the most influential non-linear term. For these subsets, we performed linear regressions and quantified variables' marginal impacts on SS. Data visualization and descriptive statistics were also used to illustrate the findings. In the following, MARS, PDP, and the linear regressions are explained.

*MARS and PDP*

To capture the non-linearity and interaction terms in the data, MARS breaks the range of the variables into bins with the use of "knots" or "cut points", each of which will be treated with separate linear models. At each knot, piecewise linear basis functions are built in the following form:

$$(x-t) = \begin{cases} x-t, & if\ x>t \\ 0, & otherwise \end{cases} \text{ and } (t-x) = \begin{cases} t-x, & if\ x<t \\ 0, & otherwise \end{cases}, \text{ with } t \text{ being the knot.} \tag{3}$$

The basis functions for each variable $X_j$ are then stored in set $C$:

$$C = \{(X_j - t), (t - X_j)\}, t \in \{x_{1j}, x_{2j}, \ldots, x_{Nj}\} \text{ and } j = 1, 2, \ldots, p, \tag{4}$$

with N being the total number of observations and p indicating the total number of variables.

The products of basis functions in $C$ that decrease the residual squared error the most are included in the model. These products are called “hinge functions”. As MARS uses products of basis functions as inputs, interactions between the variables are accounted for by the model. The MARS regression is represented by the following equation:

$$y = f(X) = \beta_0 + \sum_{m=1}^{M} \beta_m h_m(X) \quad (5)$$

where $h_m(X)$ is a hinge function, $M$ the total number of hinge functions included in the model, and where β is estimated by minimizing the residual sum-of-squares. For more details of the MARS method, readers are referred to the work of Hastie, Tibshirani, & Friedman (2009) [19].

In this paper, a regression in the form of equation (5) is conducted. The dependent variable ($y$) is the SS of models. Note that the reference model used to calculate the SS is included as the independent variables in the regression. This ensures that the impact of the reference model is accounted for when analyzing the impacts of the other variables. In this way, the SS of all types, i.e., $SS^{P}$, $SS^{PC}$, and $SS^{SP}$, can be analyzed in one model run. The other independent variables include all the variables in the database (see Table 1).

For each variable, the optimal number of knots is determined by $R^2$. A minimum increase in $R^2$ by 0.001[6] is required for an additional knot to be added to the model. After all the relevant inputs are included, input selection is conducted through backward elimination. The inputs with the least impacts on the residual sum-of-squares are excluded. Furthermore, a ten-fold cross-validation is also conducted to tune the hyper-parameters of the model (see Appendix E). The regression results in the most important variable and interaction terms for discussion.

PDP are created to assist the analysis of the interaction between variables. The plots show the marginal effect of one or a set of explanatory variables on the dependent variable while accounting for the average effect of the other variables in the model [21]. This is represented by the following equation:

---

[6] This value was suggested by the authors of MARS [20].

$$\widehat{f_S}(x_S) = E_{X_C}\left[\hat{f}(x_S, X_C)\right] = \int \hat{f}(x_S, X_C)\, dPX_C = \left(\frac{1}{n}\right) \sum_{i=1}^{n} \hat{f}(x_S, x_C^i), \quad (6)$$

where $x_S$ is the feature to be plotted, $X_C$ is the set of the other features in the model, and $x_C^i$ is the value of the $i^{th}$ feature in the set $C$.

For the purpose of interpretation we limit the number of features in the set S to two variables.

*Follow-up regression*

MARS identifies the independent variables with the highest influence on error measures. According to MARS result with more details that follow, forecast horizon is the most important variable. It divides the database into three separate groups that are roughly in line with the three horizon classifications — intra-hour, intra-day, and day-ahead. We use this result for follow-up linear regressions. Accordingly, identical linear regressions are conducted for each subset of data of a horizon group.

The dependent variable is the SS value and the independent variables are the other variables in the database. The regression is represented by the following equation:

$$SS = \beta_0 + \sum_{i=1}^{6} \beta_i CZ_i + \beta_7 Horizon + \beta_8 InputHist + \beta_9 InputMete + \quad (7)$$

$$\beta_{10} InputNWP + \beta_{11} InputST + \sum_{j=1}^{8} \beta_{j+11} ModClass_j + \sum_{k=1}^{3} \beta_{k+19} Reference_k +$$

$$\beta_{23} ResMin + \beta_{24} TestLength + \beta_{25} TrainLength + \sum_{l=1}^{2} \beta_{l+25} Type_l + \beta_{28} Year + \varepsilon,$$

where ε indicates the error and β is the coefficient of the explanatory variables.

## 3. Results and discussion

This section discusses the results from the analysis of the database in three parts, first MARS and PDP analysis, second linear regression results, and third best practices to improve SS.

### *3.1. MARS and PDP analysis*

This part presents the analysis of the non-linear relationship and interaction terms in the database based on MARS and PDP. First, the model fitting results are presented. Then, the most important factors and their impacts on SS are analyzed based on the regression results and PDP.

Table 2 presents the MARS regression results. The terms are the hinge functions used as inputs for the model, i.e., $h_m(X)$ in equation (5). The coefficients indicate the weight of the hinge function in the model, i.e., $\beta_m$ in equation (5).

Table 2: MARS regression results. The $R^2$ is 45.98% and the generalized $R^2$ is 44.39%.

| Term | Weight |
|---:|---:|
| (Intercept) | 2.130e+02 |
| h(Horizon-345) | 5.826e-01 |
| h(345-Horizon) | -6.991e-01 |
| h(345-Horizon)*h(ResMin-15) | 5.790e-04 |
| h(345-Horizon)*h(15-ResMin) | 1.121e-02 |
| h(345-Horizon)*h(2019-Year) | -3.592e-03 |
| ReferencePersistence*h(TrainLength-1095) | -4.781e-02 |
| InputMete*ReferencePersistence | 1.046e+01 |
| h(345-Horizon)*h(183-TestLength) | 2.826e-04 |
| h(Horizon-75)*ReferencePersistence | -5.504e-03 |
| h(75-Horizon)*ReferencePersistence | -3.601e-01 |
| ReferencePersistence*h(15-ResMin) | 2.483e+00 |
| ReferencePersistence*h(Year-2021) | 1.731e+01 |
| ReferencePersistence*h(2021-Year) | 1.066e+00 |
| InputNWP*ReferencePersistence | 9.222e+00 |
| h(345-Horizon)*ModClassReg | -3.499e-02 |
| CZE*ReferencePersistence | -1.562e+01 |
| ReferencePersistence*TypeSources | 1.957e+01 |
| TypeSources | -1.806e+01 |
| h(345-Horizon)*TypeSources | 5.157e-02 |
| ReferenceSP | 1.675e+01 |
| CZN | 8.348e+00 |
| ReferencePersistence*h(TrainLength-580) | 4.349e-02 |
| h(Horizon-35) | -5.842e-01 |
| ModClassReg*ReferencePersistence | -1.047e+01 |
| TypeSources*h(Year-2019) | 2.500e+00 |
| h(15-ResMin) | -2.457e+00 |
| ModClassML*ReferenceSP | -4.833e+00 |

The weight of the term $\mathrm{h}(X_j - \mathrm{t})$ or $\mathrm{h}(\mathrm{t} - X_j)$ measures the marginal effect of the variable $X_j$ on the dependent variable (SS) for $X_j > t$ or $t > X_j$, respectively. The weight of $\mathrm{h}(X_i - \mathrm{a}) * \mathrm{h}(X_k - \mathrm{b})$ measures the marginal effect of the interaction term of the two variables $X_i$ and $X_k$, when $X_i > a$ and $X_k > b$. Likewise, the weights of $\mathrm{h}(a - X_i) * \mathrm{h}(X_k - \mathrm{b})$, $\mathrm{h}(X_i - \mathrm{a}) * \mathrm{h}(\mathrm{b} - X_k)$, and $\mathrm{h}(a - X_i) * \mathrm{h}(\mathrm{b} - X_k)$ measure the marginal effect of interaction terms between $X_i$ and $X_k$, when $a > X_i$ and $X_k > b$, $X_i > a$ and $b > X_k$, and $a > X_i$ and $b > X_k$, respectively. For more details on interpreting MARS results, readers are referred to the work of Hastie, Tibshirani, & Friedman (2009) [19]. Due to the complexity in interpreting the weights of the terms in the MARS regression, the analysis of MARS results will be mainly done through the partial dependence plots (PDP), which visualize the effect of a set of variable on SS that accounts for the effects of other variables in MARS regression.

A key result can be derived from Table 2: forecast horizon is the most important variable and interacts with most of the other variables in the regression. Eleven out of 28 interaction terms include the forecast horizon variable. Furthermore, the cut points of horizon are noteworthy. Distinct interactions are observed for horizon > 345 minutes, 75–345 minutes, and < 75 minutes. Note that this aligns well with the definition of intra-hour forecasts (Horizon <= 60 minutes), intra-day (60 minutes< Horizon <= 360 minutes) and day-ahead forecasts (Horizon > 360 minutes) in many studies (e.g., [3]). This reveals that intra-hour, intra-day and day-ahead PV forecasts are separate sub-disciplines in solar forecasting, with different parameter weights explaining forecast quality. This motivates our follow-up analysis using linear regression for each horizon classification in the next sub-section.

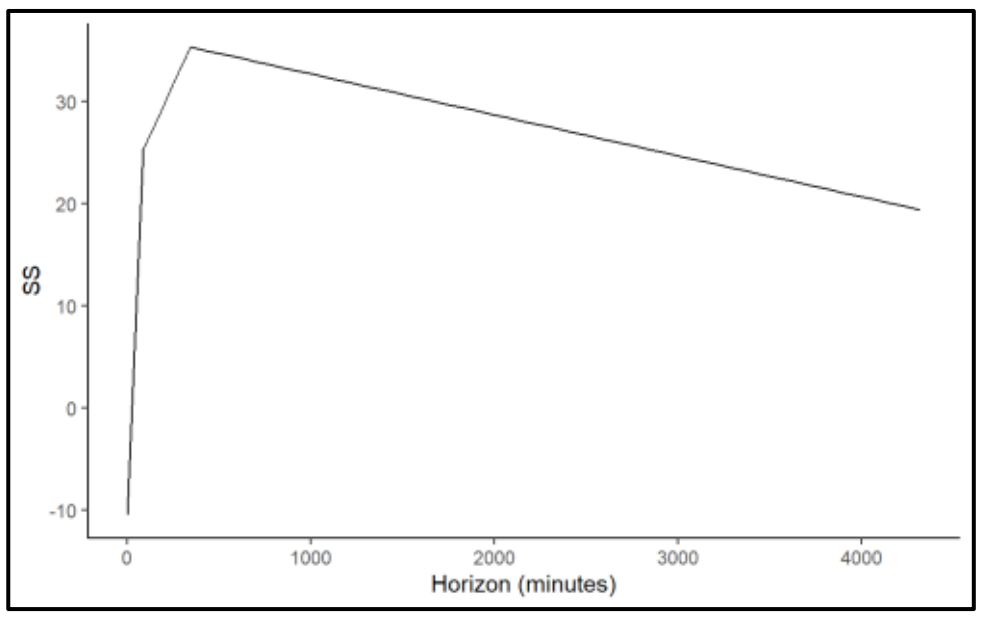

Figure 3: PDP for horizon variable. The figure shows the partial dependence of SS on horizon, accounting for the effects of all the other variables in the MARS regression. Non-linearities are observed and distinct pattern can be seen for horizon < 75 minutes, 75–345 minutes, and > 345 minutes. These cut points approximate the thresholds for intra-hour, intra-day, and day-ahead forecasts, respectively. Note that PDP plots visualize the MARS regression results, which already controls for the impacts of reference models. Therefore, the skill score value can be interpreted irrespective of reference models. This applies to all other PDPs in this paper.

The PDP in Figure 3 illustrates how MARS endogenously groups horizons threefold. As can be seen, the slope of the line is particularly sharp and positive for horizon < 75 minutes, still positive but less steep for horizon 75–345 minutes, and then turns negative for horizon > 345 minutes. This indicates that for low values of horizon (~ < 345 minutes), SS increases with horizon, which indicates that forecasts show more accuracy for higher horizon. This might seem contradictory with many previous studies showing that forecast accuracy decreases with horizon [22]. However, note that the metric used in this meta-analysis is SS, which shows a relative accuracy of forecasts compared to reference models. Therefore, an increase in SS can come from forecasts' better performance and/or reference models' worse performance. As horizon increases, the reference models show inefficiency in capturing future dynamics [12]. As a consequence, despite possible higher absolute errors (e.g., RMSE), the forecasts are expected to perform (relatively) better for longer horizons. Notably, this only holds for low values of horizon. For horizon > 345 minutes, forecast accuracy decreases with horizon, possibly due to the large increase in the absolute errors. This shows the challenges in predicting power values as horizon gets larger.

PDP are also created for other numerical variables in the MARS regression for easy interpretation of the results from Table 2. Figure 4 presents the plots for single variables and Figure 5 includes the plots for interaction terms between two variables.

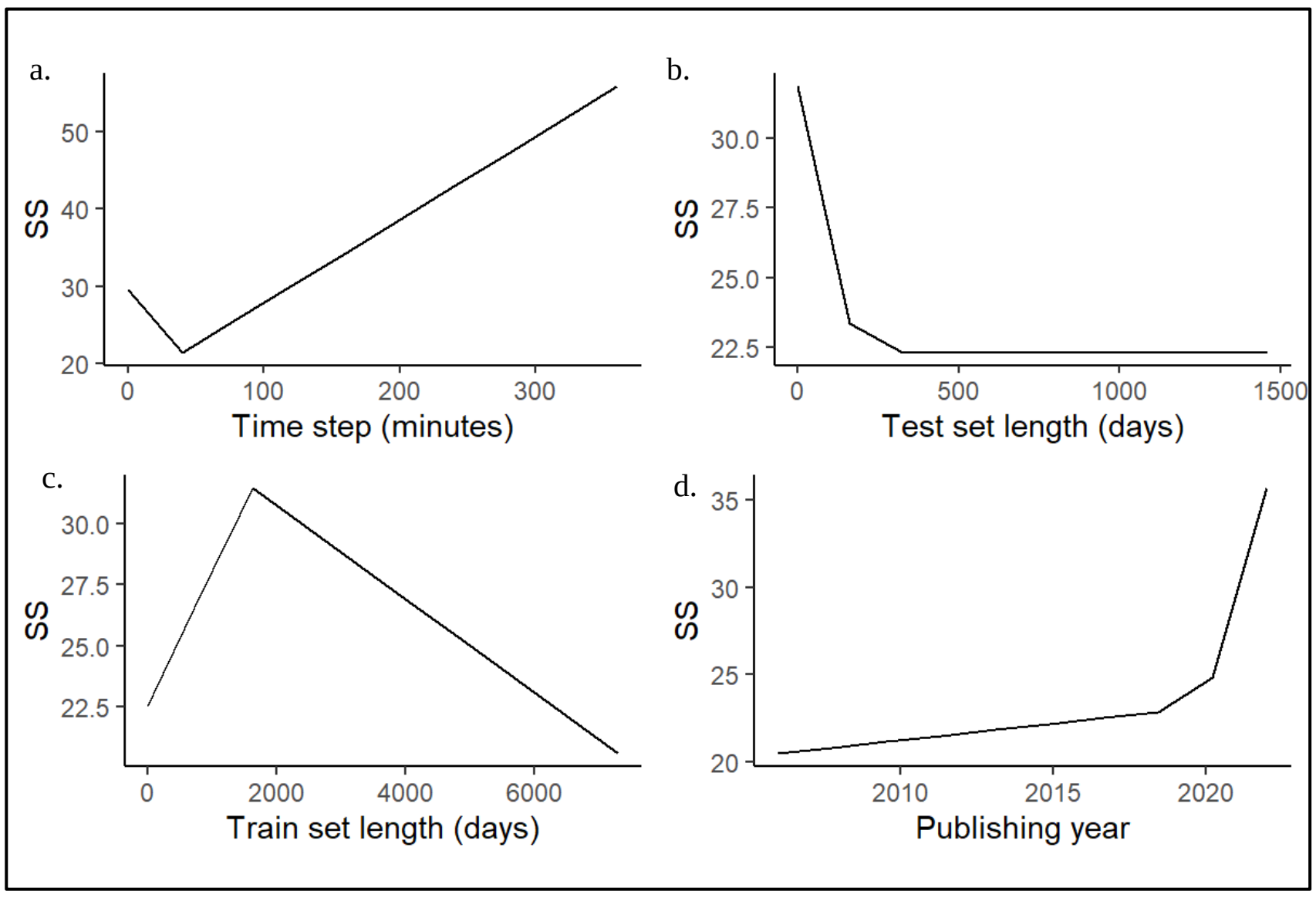

Figure 4: PDP for single numerical variables. The figure shows the partial dependence of SS on each variable, accounting for the effects of all the other variables in the MARS regression.

Figure 4 illustrates the non-linearity in the relationship between SS and variables that have been identified as being vital by MARS. Figure 4a indicates that for a time step > 30 minutes, SS increases with the length of the time step, which implies that SS is higher for lower resolution (see Appendix A for the definitions of time step and forecast resolution). This confirms the fact that to forecast power at a low resolution (e.g., half-daily or daily) can achieve higher accuracy than to forecast at a higher resolution (e.g., hourly or half-hourly). Noteworthily, higher SS is also observed for very small time steps (e.g., every 5 or 10 minutes). This is possibly due to a lower level of volatility in the weather conditions for very small time steps, leading to less uncertainty and more forecast accuracy.

Figure 4b illustrates a negative correlation between SS and test set length. The correlation is particularly high for short test sets (less than180 days) and becomes insignificant for long test sets (at least 365 days). This aligns with previous studies [5] that test sets of at least one year can better validate models and avoid "cherry picking" in reporting errors.

Figure 4c shows that a longer train set can enhance SS values. Certainly, models trained with more data can see more patterns and therefore predict better. However, for already large train set (e.g., around 2000 days), adding more training data can lead to a decrease in forecast accuracy. This can come from the "over-fitting", where models try so hard to fit the training data that it fails to generalize in the validation data.

Figure 4d illustrates a remarkable progress in scientific achievements in enhancing forecast accuracy in solar forecast domain. An increasingly positive correlation between SS and publishing year of papers is observed. It took thirteen years (from 2006 to 2019) to increase SS by 3 percentage point (pp) but only one year (from 2021 to 2022) to achieve 10-pp improvement in SS.

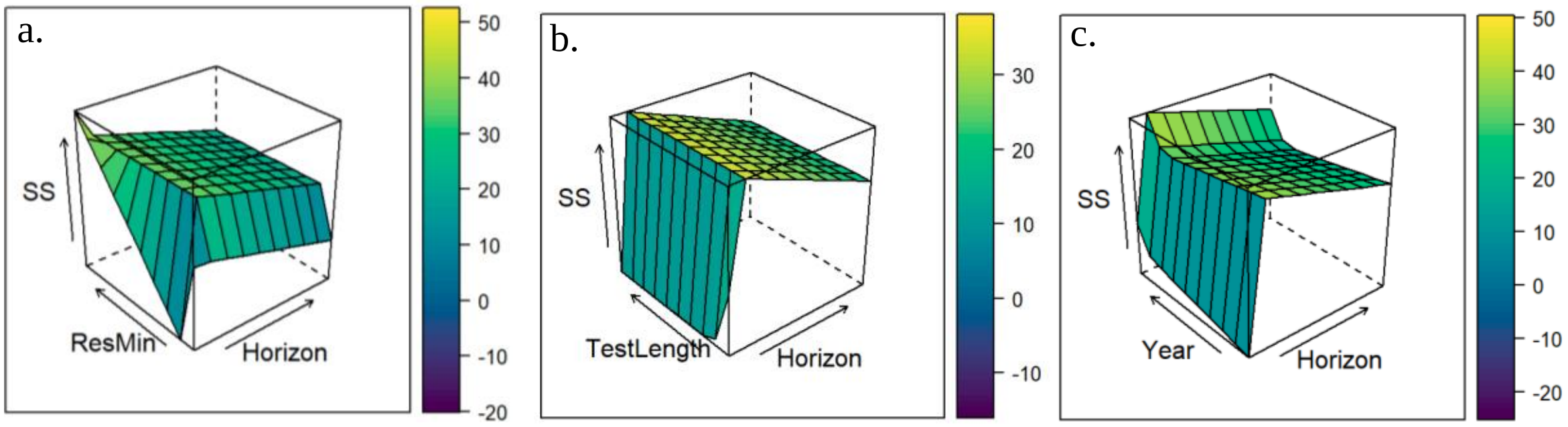

Figure 5: PDP for two-degree interactions of numerical variables. The figure illustrates the partial dependence of SS on the interaction of different numerical variables. The direction of the arrow indicates the increasing direction of the value. The color represents the value level of SS.

Figure 5 shows the partial dependence of SS on interaction terms of numerical variables that have been captured in MARS. Resolution and horizon's marginal effect on SS is presented in Figure 5a. SS is observed highest for short horizons and high values of resolution length. For very large horizon, SS is lower, and the impact of resolution seems insignificant.

The interaction between test set length and horizon is pictured in Figure 5b. Higher SS is observed for shorter horizons while the change in SS with test set length for each horizon level is hard to see by the change of the colour. As can be seen, SS seems more affected by horizon than test set length.

Figure 5c describes the marginal effects of publishing year and horizon on SS. Highest SS is observed for recent years with short horizons. Furthermore, the gap of SS between short and long horizons is more significant in recent years than several years before. This implies that much progress has been done in improving forecast accuracy, particularly for short horizon forecasts.

Overall, different effects of variables on SS have been detected using MARS. It can be learned that these effects can sometimes be non-linear, and interaction between variables also matters. By accounting for both non-linearity and interaction terms in the regression, our findings represent more accurate the marginal impacts of variables on SS. Furthermore, we show that horizon variable has a leading impact on SS and dominates in the interaction terms. This motivates separate analysis for each horizon class and leads to the next sub-section.

*3.2. Follow-up regression analysis for different horizons*

In this section, a linear regression according to equation (7) is conducted for intra-hour (<= 1 hour), intra-day (1–6 hours), and day-ahead (> 6 hours) horizons. Table 3 presents the regression results.

Table 3: Linear regression results.

| | *Dependent variable:* SS | | |
|---|---|---|---|
| | Intra-hour (1) | Intra-day (2) | Day-ahead (3) |
| CZA | 5.013*** | 6.553*** | 5.988** |
| | (1.719) | (1.457) | (2.498) |
| CZB | 3.783*** | 2.058** | 2.722* |
| | (1.226) | (0.961) | (1.630) |
| CZD | 7.984*** | -2.599 | 1.638 |
| | (1.324) | (1.687) | (1.515) |
| CZE | -19.940*** | | -8.358** |
| | (3.158) | | (3.262) |
| CZNA | 0.322 | 7.339*** | 7.052* |
| | (2.888) | (1.453) | (3.753) |
| Horizon | 0.334*** | 0.060*** | -0.004*** |
| | (0.043) | (0.004) | (0.001) |
| InputHist | 5.672** | 8.194*** | -6.426*** |
| | (2.421) | (1.285) | (2.086) |
| InputMete | -0.441 | 9.084*** | 5.055*** |
| | (1.169) | (1.005) | (1.305) |
| InputNWP | -9.037*** | -2.284* | 11.575*** |
| | (2.394) | (1.209) | (1.354) |
| InputST | 3.598** | 3.941*** | -5.511*** |
| | (1.466) | (1.427) | (1.413) |
| ModClassEns | 0.088 | -6.951*** | 8.327*** |
| | (2.372) | (1.908) | (1.854) |
| ModClassEns_Hyb | 12.837*** | 21.213*** | 7.034*** |
| | (4.188) | (6.395) | (1.559) |
| ModClassHybrid | 8.605*** | -19.317*** | -11.311*** |
| | (2.612) | (3.052) | (2.455) |
| ModClassImageBased | 1.164 | 10.292*** | -0.212 |
| | (2.157) | (1.546) | (1.778) |
| ModClassML | -3.134* | -0.117 | -1.195 |
| | (1.680) | (1.064) | (1.570) |
| ModClassNWP | -13.938 | -17.391** | -14.262*** |
| | (20.808) | (7.171) | (3.147) |
| ModClassReg | -11.048*** | -5.267*** | -0.555 |
| | (2.245) | (1.587) | (3.007) |
| ReferencePersistence | 15.936*** | 51.517*** | 10.677*** |
| | (3.464) | (3.671) | (2.416) |
| ReferenceSP | 10.731*** | 27.267*** | 9.026*** |
| | (3.433) | (3.452) | (2.576) |
| ResMin | 0.020 | 0.056*** | 0.009 |
| | (0.037) | (0.021) | (0.008) |
| TestLength | -0.011*** | -0.008*** | 0.018*** |
| | (0.002) | (0.003) | (0.003) |
| TrainLength | -0.001*** | 0.002*** | 0.004*** |
| | (0.0005) | (0.0004) | (0.001) |
| TypeSources | 2.177 | -17.178*** | 5.055*** |
| | (1.825) | (1.518) | (1.957) |
| Year | 1.795*** | 1.493*** | 1.160*** |
| | (0.254) | (0.201) | (0.324) |
| Constant | -3,637.375*** | -3,031.775*** | -2,322.335*** |
| | (512.621) | (407.525) | (653.369) |
| Observations | 2,116 | 1,555 | 1,016 |
| $R^2$ | 0.213 | 0.513 | 0.456 |
| Adjusted $R^2$ | 0.204 | 0.506 | 0.443 |
| *Note:* | | | *: p-value < 0.1, **: p-value <0.05, ***: p<0.01 |

Columns (1) through (3) represent intra-hour, intra-day, and day-ahead, respectively. The coefficient [$(\beta)$ in the Equations (7)] of continuous variables shows how SS changes given one unit change of the variable. Similar to MARS, the linear regressions also control for the impacts of reference models, allowing SS to be interpreted regardless of the choice of reference models. The coefficient of categorical variables indicates the gap of SS between the category of interest and the baseline category. The coefficient of dummy variables measures the change in SS when the variable equals 1. The standard error is shown in parentheses under the coefficient and represents the average distance between the observed values and the regression line. The coefficient is considered statistically significant if the p-value is lower than 0.1 so that the null hypothesis that the variable has no correlation with the dependent variable can be rejected. If p-value is lower than 0.01, the variable is considered highly significant in the regression. The "Constant" reported toward the end of the table indicates the average error when all explanatory variables are set to zero [$(\beta_0)$ in the Equations (7)]. Furthermore, $R^2$ and adjusted $R^2$ are also reported in the table to show the explanatory power of the regressions.

In the following sub-sections, the effect of each variable on SS is discussed.

#### *3.2.1. Climate zone (CZ)*

The first block of Table 3 shows the coefficients of CZ variables, including CZ of A, B, D, E, and NA (not available information of climate zone) types. Climate type C is set as the baseline variable, as it covers a majority of the database (see Figure 2). Our analysis reveals another key result: forecast quality is highly dependent on climate zone.

As discussed, the SS metric is believed to effectively reduce the impacts of location and climate conditions thanks to the relative measurement. However, the regression shows that the location-contingent variable such as CZ still have potential impacts on SS. Most of the coefficients are highly statistically significant. Except for zone E, the other zones show relatively higher SS than zone C. It can be inferred that better forecasts are more commonly observed in zones A, B, and D than in zones C and E. This is relatively opposite to the findings from X. Yang, Yang, and Wang (2021) [9], which show that higher RMSE is found for zones A and D. One explanation for this is that zones A and D are more difficult to forecast, resulting in higher absolute error values. The performance gap between forecast models and the reference models becomes more obvious when the forecasting becomes more difficult.

This leads to a higher SS value for these zones. More importantly, some climate types having better forecasts than the others indicates a challenge in transferring insights from one climate zone (e.g., based on the many good papers published for the USA) to other climate zones (e.g., zones where significant growth in PV capacity is expected in the near future, such as Northern Africa or the Middle East, but not yet studied). If possible, a meta-analysis should be implemented to correct for climate zone and other effects, even when the SS metric is used.

#### *3.2.2. Horizon*

The coefficients for horizon variable in Table 3 show that the correlation between horizon and forecast accuracy is significantly positive for intra-hour and intra-day forecasts (+0.334 and +0.06, respectively). However, the correlation turns negative for day-ahead forecasts (-0.004). This is identical to the non-linearity observed for the horizon variable, as discussed above.

#### *3.2.3. Input*

Keeping examining the results from Table 3, we show that historical data (InputHist) is more effective for intra-hour and intra-day than for day-ahead forecasts. Using historical data as among inputs can increase SS by 5.67 pp for intra-hour and 8.19 pp for intra-day forecasts. In the meanwhile, locally measured meteorological data (InputMete) and numerical weather predictions data (InputNWP) are more helpful for intra-day and day-ahead forecasts than for intra-hour forecasts. Having InputMete can help models enhance SS by 9 pp for intra-day while InputNWP corresponds to nearly 12 pp of SS increase for day-ahead forecasts. Notably, using NWP data can lead to less accurate forecasts for intra-hour by around 9 pp. This is possibly because the default temporal resolution of NWP data is often every 6 hours to 12 hours, which is not helpful for intra-hour forecasting [12]. Furthermore, the spatial temporal information of the power plants and their neighbors (InputST) also contributes to improving forecast accuracy. The regression result shows that using InputST shows SS higher by around 4 pp for intra-hour and intra-day forecasts. Noteworthily, InputST is less effective for longer horizons such as day-ahead. This also aligns with previous findings [23] that skill score seems independent of network layout for longer horizons.

Regarding the image-based inputs such as sky and satellite images, these inputs are inherent within the image-based models. Therefore, their effects on SS will be discussed together with model variables.

#### 3.2.4. Models (ModClass)

The next seven variables of Table 3 present different forecast model classifications. The baseline is the time series method (TS). The coefficients thus show how each model class performs relative to TS. As observed, ensemble models (Ens) achieve higher forecast accuracy by 8.5 pp than TS for day-ahead forecasts. Ensemble–hybrid methods (Ens_Hyb) have the best performance, outperforming TS by around 13 pp for intra-day, 21 pp for intra-hour, and 7 pp for day-ahead horizon. Hybrid models outperform TS by 8.6 pp for intra-hour forecasts. Image-based methods show efficiency for intra-day horizon, improving SS by over 10 pp compared to TS. This indicates a significant contribution of image-based inputs to enhancing forecast accuracy especially for intra-day forecasts. The other methods, including individual machine learning (ML) models, NWP models, and regressions, do not show superiority to TS models. Overall, combining methods and forecasts generate higher accuracy. However, TS methods still have very good performance compared to more complex methods such as individual ML models. Considering the simplicity and thus less computational burden of TS, attempts should be made to improve these models before starting with more complex methods.

#### 3.2.5. Reference model

As discussed, the reference models used to calculate SS are included as an independent variable so that their effects on SS can be controlled for in the regression. The baseline variable is $SS^{CP}$, i.e., forecasts benchmarked against a convex combination of smart persistence and climatology models. Therefore, the coefficients indicate the change in SS when changing from $SS^{CP}$ to $SS^{P}$ or $SS^{SP}$. As can be seen, $SS^{P}$ is higher than $SS^{CP}$ by roughly 11–52 pp. The gap is highest for intra-day horizon and lowest for day-ahead horizon. $SS^{SP}$ is higher than $SS^{CP}$ by 9–27 pp. Most of the coefficients are highly statistically significant at p-value < 0.01. This confirms the suggestion of many scholars that the choice of reference model can have a significant impact on SS level. $SS^{CP}$ is generally much lower than the other two SS types.

#### 3.2.6. Resolution (ResMin)

In terms of the resolution variable, the regression results (Table 3) show that higher SS is observed for longer time step (which means lower resolution). This is highly statistically significant for intra-day forecasts, possibly due to a larger range of resolution values for intra-day (1–360 minutes) than for the intra-hour (1–60 minutes) and day-ahead (15–360 minutes) in the database.

#### *3.2.7. Test set length (TestLength)*

All coefficients of test set length (Table 3) are highly statistically significant. A negative correlation between this variable and SS is observed for intra-hour and intra-day forecasts, confirming previous studies' results [5]. For day-ahead forecasts, however, the correlation turns positive. This is possibly because reference models show more inefficiency for long horizons, which becomes worse on long test sets, leading to higher SS for forecasts.

#### *3.2.8. Train set length (TrainLength)*

Train set length also shows statistically significant correlation with SS. As can be seen from Table 3, an additional day in the train sets can improve SS for intra-day and day-ahead forecasts by 0.002 and 0.004 pp, respectively.

#### *3.2.9. Type of forecasts*

The type of forecasts, i.e., either PV or solar resource forecasts, is included as a dependent variable to control for any possible effects from different forecast targets. The baseline is PV forecasting. The coefficients (Table 3) measure how much higher SS is for solar resource forecasting compared with PV. They show that solar resource forecasts achieve lower SS for intra-day but higher SS for day-ahead. The coefficient for intra-hour is insignificant. Therefore, no forecast type shows superiority than the other.

#### *3.2.10. Year of publication*

The scientific progress in the field is represented by the change in SS with the publication year of papers. The regression results indicate an annual increase in SS by 1.2–1.8 pp. The largest progress is observed for intra-hour forecasts, followed by intra-day, and day-ahead. This is identical to the partial dependence of SS on year variable observed in Figure 4e.

### *3.3. The best practices in solar forecasting*

Based on the in-depth analysis in the previous sections, a summary on the best practices in solar forecasting is provided.

#### *3.3.1. Forecast verification*

SS is based on a relative measurement approach. Therefore, it can significantly reduce the impact that differences between data sets will have on forecast accuracy. However, some impacts remain and should be accounted for when comparing forecasts. A meta-analysis controls for these impacts and generates insights that can be applied globally.

We recommend future work report at least the SS based on RMSE as RMSE enables inter-model comparison with the majority of previous studies. We also show that the choice of reference model to calculate SS can significantly influence the values of SS. Therefore, we suggest a consensus on using a convex combination of smart persistence and climatology as the reference model. The convex combination of smart persistence and climatology models achieves a higher forecast accuracy than simple persistence or smart persistence [24] and is thus a more ambitious benchmark.

#### *3.3.2. How to improve forecast accuracy?*

Our analysis shows that the effects of variables on SS or forecast accuracy is complex, which can include non-linearity and variables' interactions. This emphasizes that generalizing findings between studies and transferring knowledge between regions must be conducted very carefully and account for all the discussed effects. Through the meta-analysis in this paper, some conclusions can be applied globally as follows.

First, lower resolutions and shorter horizons can lead to higher SS. Second, more training data can improve models' performance. However, the amount of training data should be controlled so that there is no over-fitting. The database shows that the training data of around 2000 days achieve the highest forecast accuracy.

Third, regarding the usage of inputs, the horizon is particularly relevant. For intra-hour, historical data and spatial temporal information of power system and its neighbors are highly helpful. For intra-day, image-based inputs such as sky and satellite images show the most important role, followed by locally measured meteorological data, historical data, and spatial temporal information. For day-ahead, NWP variables improves models the most. Locally measured meteorological data is also very efficient here.

Next, the choice of forecast model is also important. Generally, for all horizons, ensemble–hybrid models show the best performance, and thus are recommended for solar forecasting. However, considering the accuracy–complexity trade-off, it is also recommended to bear in mind the pragmatic use of simple methods such as time series – potentially optimizing them with techniques on data pre- and post-processing – before starting with more complex methods.

## 4. Conclusions

This paper provides the first meta-analysis of solar forecasting based on the skill score metric. A comprehensive search has been conducted on Google Scholar for all literature in the field published from 2006 to 2022. A total of 2,335 search results have been screened, and a database of 4,687 observations from the solar forecasting literature with 11 variables has been built. This database is large enough to control for various factors and produce robust, statistically significant results that can be applied globally. The key take-aways can be summarized differentiated by non-linear results derived from the multivariate adaptive regression spline modelling (MARS) and partial dependence plot (PDP) as well as by the linear regression for sub-groups identified by MARS:

First, MARS revealed two key insights on non-linearity and complex interaction terms in the database:

- The model identified forecast horizon class as a significant non-linearity. Based on this result, a key take-away is that what works for intra-hour forecasts does not necessarily work for intra-day or day-ahead forecasts. Therefore, the analysis of solar forecasts should be done separately for each forecast horizon.
- There have been substantial improvements in solar forecast accuracy over time, especially in recent years. These improvements are higher for intra-hour and intra-day than for day-ahead forecasts. This is good news both for researchers in the field who have made this development possible as well as for energy system planning, as solar power will become easier to integrate into electricity systems with better forecasts.

Second, a follow-up linear regression based on the MARS and PDP analysis quantified the marginal impact of important variables on skill score. Key findings are:

- Location-related variables such as climate zones show statistically significant correlation with SS. Hence, the relative measurement approach of SS is not sufficient to allow a direct comparison of forecasts. A meta-analysis that can account for different effects is important for knowledge transfer between regions. This result is important for future research, in particular as past research has mostly focused on developed countries (e.g. the US), while developing countries (which often have great solar potential), have hardly been researched. First, researchers should be careful when transferring insights on optimality of methodologies from one climate zone to another. Second, front runner papers on new regions and climate zones should not necessarily be expected to already outperform SS derived from studies in better researched zones.
- The linear regression also added to the key result from MARS that intra-hour, intra-day and day-ahead forecasting are different sub-disciplines of solar forecasting. Our follow-up regression confirmed that input usage should depend on horizon class. For intra-hour, historical data and spatial temporal information of power system and its neighbors are highly helpful. For intra-day, sky and satellite images show the most important role, which can be combined with any other inputs. For day-ahead, NWP variables and locally measured meteorological data are very efficient.
- Regarding the inter-model comparison, ensemble–hybrid models achieve the most accurate forecasts. Many methods do not show a robust superiority to time series methods. Therefore, it is recommended to consider improving the performance of simple models through using different techniques and processing input data before moving on to complex models.

These findings provide important guidance for future solar power forecasters to improve forecast accuracy. Highly applicable recommendations regarding the forecast methodologies and input usage are made separately for each forecast horizon. This is crucially important for regions where a tremendous growth in solar power generation is expected, but the research on solar forecasting is not as robust. For example, Northern Africa and the Middle East might enhance their investments in solar power in the medium-term, considering their very good solar irradiation conditions [25]. Transferring knowledge gained from previous studies to these regions can provide a significant benefit to forecasting.

There are other important factors that have not been analyzed in this paper due to a lack of observations. Variables such as the quality of the cameras, the type of installation (e.g., non-tracking, 1- or 2-axis tracking), the lead-time, and the physical size of power plants are also particularly important to the forecast accuracy and should be included in future analysis.

**Data Availability**

The database used for all results presented in this paper is provided with this paper as Supplementary Data. The database is also publicly available in the ZENODO repository, DOI: 10.5281/zenodo.7274381 (https://doi.org/10.48550/arXiv.2208.10536).

**Acknowledgements**

The authors gratefully acknowledge the financial support by the BMBF project Energie-Innovationszentrum (EIZ), project number 03SF0693A.

## Appendix A. Background for data extraction – independent variables explained

Through our review of historical studies and literature surveys on solar forecasting (see Table A.1, we show that the top ten important variables that significantly influence forecast accuracy are (alphabetically) climate conditions, forecast horizon, inputs, models of forecasts, reference model (used to calculate skill score), resolution of forecasts, test set length, train set length, the type of forecasts (solar resource or PV output forecasts), and the year of publication of papers.

Table A.1: Literature surveys on solar forecasting

| No | Year | Ref. | Summary |
|---|---|---|---|
| 1 | 2022 | [26] | Review of solar and wind forecasting. Provides detailed discussion of forecasting methodologies and covers both point and interval forecasting. Forecasting of ramping events is also discussed. |
| 2 | 2022 | [27] | Review of PV and solar resource forecasting. Gives an overview of the status of methodologies and different forecast goals (i.e., PV, solar irradiance), and emphasizes the importance of hybrid methods and optimization techniques. |
| 3 | 2022 | [28] | Review of very short-term forecasting (minutes to hours ahead) for wind and solar. Discusses an open-source case study and suggests the best practice for forecast evaluation and research. |
| 4 | 2021 | [29] | Review of wind and solar forecasting with a focus on deep learning models. Discusses distribution of research in terms of locations and methodologies. Discussion of probabilistic and multistep-ahead forecasting methods also provided. |
| 5 | 2021 | [14] | Review of progress in intra-hour irradiance forecasting methodologies. Discusses the theories behind methodologies and different techniques and their application. |
| 6 | 2021 | [30] | Review of artificial intelligence in solar PV systems. Uses text mining to collect, analyze, and categorize papers on solar forecasting, detection, design optimization, and optimal control. |
| 7 | 2021 | [31] | Review focusing on current advances and challenges in PV forecasting. Discusses available techniques, especially machine learning (ML), ensemble, and hybrid forecasting performance. |
| 8 | 2021 | [32] | Review of solar forecasts. Summaries optimization methods and inputs for models. Comparison between probabilistic and deterministic forecasting also discussed. |
| 9 | 2021 | [33] | Review focusing on deep learning models for solar irradiance forecasting. Discusses the factors influencing forecast accuracy and basic structure of deep learning. Also compares selected models. |
| 10 | 2021 | [34] | Review of deep learning models for PV forecasting. Highlights superiority of deep learning and discusses potential application and possible future challenges. |
| 11 | 2021 | [35] | Review of artificial neural network (ANN) and fuzzy system techniques for solar radiation forecasting. Discusses and compares methodologies for different inputs. |
| 12 | 2021 | [36] | Review of intra-hour solar forecasting using sky images. Discusses the most important sky image features to improve forecast accuracy. |
| 13 | 2021 | [37] | Review focusing on hybrid models for solar radiation forecasting. Discusses and compares models regarding their application and potentials for each forecasting horizon. |
| 14 | 2021 | [38] | Review of solar irradiance and power forecasting that provides a thorough classification of inputs, methods and techniques. |
| 15 | 2021 | [39] | Review of operational solar forecasting. Discusses four key technical aspects: (1) gauging the goodness of forecasts, (2) quantifying predictability, (3) forecast downscaling, and (4) hierarchical forecasting. Analysis of a case study is also provided. |
| 16 | 2021 | [40] | Review focusing on measures to validate forecasts. Discusses different aspects of a good forecast and proposes new measure of predictability of solar irradiance. |
| 17 | 2021 | [9] | Proposes a global database of reference for solar forecast accuracy. Also provides guidelines for calculating errors so models can be easily comparable. |
| 18 | 2020 | [3] | Review of short-term PV output forecasts. Many important variables such as forecast horizon and data processing techniques are discussed. |
| 19 | 2020 | [41] | Review of short-term PV output forecasts discussing basic principles, standards, and different methodologies. |
| 20 | 2020 | [6] | Review of PV forecasting proposing new way of classifying models based on forecast horizon. Also discusses inputs, outputs, forecasting methodology and performance metrics. |
| 21 | 2020 | [42] | Review of highly advanced methods for PV output forecasting, especially recent developments in ML, deep learning (DL), and hybrid methods. |
| 22 | 2020 | [43] | Review of both solar irradiance and PV output forecasting, focusing only on ANN models. Highlights superiority of ANN hybrid models and emphasizes importance of data input quality and weather classification. |
| 23 | 2020 | [44] | Review focusing on standardizing forecast verification approaches in deterministic solar forecasting. Discusses practical issues during verification and suggests best practice in forecast verification. |
| 24 | 2019 | [45] | Review of ML and hybrid methods for solar irradiance and PV output forecasts that suggests the superiority of ML-based hybrid models. |
| 25 | 2019 | [46] | Review of solar energy nowcasting. Provides an overview of applications, solar resource data, evaluation procedures, modelling methods, and emerging technologies in solar nowcasting. |
| 26 | 2019 | [47] | Guideline for best solar forecasting practices. Provides a set of characteristics to facilitate comparison, comprehension, and communication within the solar forecasting field. |
| 27 | 2019 | [48] | Guideline for the reference model used to calculate the skill score metric. Climatology, persistence, and their linear combination are discussed. An empirical case study is also conducted. |
| 28 | 2018 | [49] | Review of the development of PV output forecasts and model optimization techniques. Suggests that ANN and support vector machine (SVM) models have accurate and robust performance. |
| 29 | 2018 | [50] | Review of PV output forecast methods indicating superiority of ANN and SVM models. Also suggests ensemble methods have much potential in enhancing forecast accuracy. |
| 30 | 2018 | [11] | Detailed analysis of key aspects of solar forecasting using text mining. Good source of terminology explanation especially for beginners to the field. |
| 31 | 2017 | [51] | Review of very short-term PV output forecasts with cloud modelling. Suggests that hybrid models, combining physical with statistical models, can enhance forecast accuracy, especially when PV outputs have rapid fluctuations. |
| 32 | 2017 | [52] | Review of solar radiation forecasting using ANN models. Discusses inputs and selected models, and also compares ANN with conventional regression models and indicates the superiority of ANN. |
| 33 | 2017 | [53] | Review of forecasting methods of solar irradiation using ML approaches. Discusses and compares different ML models. |
| 34 | 2016 | [54] | Review of PV output forecasts. Discusses key features of PV forecasting and suggests the dominance of ML-based models. |
| 35 | 2016 | [55] | Discussion of ML and classical methods for PV output forecasting that supports the use of ML models and data processing techniques. |
| 36 | 2013 | [8] | Review of forecast evaluation metrics. Proposes new metric that directly compares forecasting error with solar variability. An empirical case study is conducted and models are compared. |
| 37 | 2008 | [56] | First review of ANN-based models for PV output forecasting. Suggests a high potential for ML techniques in enhancing forecast accuracy. |

The variables are discussed in details as follow.

*Climate zone*

The generation of solar power depends strongly on climatic conditions [9]. The Köppen-Geiger (KG) classification of climate zones is the most frequently applied in many fields [57] and has been widely applied in solar forecasting [10]. The KG system divides terrestrial climates into five major types represented by capital letters: A (equatorial), B (arid), C (warm temperature), D (snow), and E (polar). Within each major type, the climates are further split based on the precipitation level and temperature. In this paper, we classify regions' climate zones based on the major type only.

Previous studies have shown that RMSE values tend to be smaller for climate types B and E than for types A and D [9]. Our paper will analyze whether this result holds for SS. To do this, information on the location of the forecast is used to generate longitude and latitude using Google Geocoding [58]. KG climate zones are then identified based on regions' longitude and latitude and recorded under a categorical variable "CZ".

*Forecast horizon*

The (forecast) horizon measures the time that the forecast looks ahead [49]; it is the period between the moment the forecast is made and some specified time in the future. Forecasts are less accurate as the horizon increases [10].

For the database, the horizon information measured in minutes is extracted from papers and stored in a numerical variable named “Horizon”.

*Forecast Input*

The use of input data significantly influence forecast accuracy [14]. The most direct input for solar forecasting is historical data on solar power infeed. Locally measured meteorological data and numerical weather predictions (NWP), as well as data derived from sky and satellite images are also very important [59] and often used as input. Spatial-temporal information of the power system and its neighbors is also believed to improve the forecast accuracy [60].

In the database, the input information is represented by dummy variables instead of categorical variables because one forecast model can use multiple inputs. The dummy variable’s value of 1 indicates a positive and 0 indicates a negative usage of the input. “InputHist” stands for historical data, “InputMete” for locally measured meteorological data, “InputNWP” for the NWP data, and “InputST” for spatial-temporal information. In terms of input from sky and satellite images, they are inherent within “image-based” model classification. Therefore, there are no dummy variables for these inputs. The impacts of these inputs will be examined through the image-based model class.

*Forecast models*

It is always of interest to know which methodology or model can achieve higher forecast accuracy. Due to a wide variety of models currently used to forecast solar power, it is better to first classify models into groups.

This paper uses the model classification which in most part closely follows D. Yang et al. (2018) [11]. Individual models are classified into ‘Time Series’ (TS), ‘Regression’, ‘NWP’, ‘ML’, and ‘Image-Based’. Furthermore, we also add three additional categories for the methods combining different individual methods: ‘Ensemble’, ‘Hybrid’, and ‘Ensemble–Hybrid’. Each model class is defined as follows:

TS method deals with a sequence of observations over successive points in time [61]. This method includes autoregressive integrated moving average, exponential smoothing, and generalized autoregressive conditional heteroskedasticity. Due to the strong impact of cloud movements and weather conditions on solar power, the multivariate models such as the autoregressive with exogenous input and vector autoregressive models are believed to have higher forecast accuracy [11].

Regression method estimates the relationship between inputs and the output and is often used to model exogenous variables. Although similarity can be found between autoregressive and regression models, classifying autoregressive in the regression category is still under argument [11]. There are many versions of regression models, ranging from linear to non-linear regression. Generalized regression models such as the least absolute shrinkage and selection operator and elastic net are also classified under this category.

NWP models provide outputs of weather variables, including the solar irradiance indicators (mostly GHI and, to a lesser extent, DNI). Often, scholars combine the outputs from different NWP models and/or use post-processing techniques to improve the forecasts from NWP. Some of the most popular NWP models are GEM [62], GFS [63], IFS [64], and NAM [62].

ML methods learn patterns and optimize model parameters from the data at the same time. Based on this framework, ML can predict future solar power based on historical observations. ML is also very helpful in extracting features and processing sky images. Therefore, ML models have been widely applied in solar power forecasting [53]. Artificial neural network and support vector machine/regression are the most popular ML methods observed in the publications. Other methods such as k-nearest neighbours, random forest, and gradient-boosted regression are also increasingly used [11].

Image-based forecasting mostly uses sky or earth imagery to extract cloud information. This information is then used to calculate solar irradiance at the surface at the time of the image and predict future cloud motion [65]. A block matching method to determine cloud motion vector [66], cross-correlation method, or particle image velocimetry are some examples of techniques used to predict future GHI based on cloud motion and current GHI values [11].

Ensemble methods combine forecasts and improve the forecast quality based on data diversity and parameter diversity [67]. To conduct an ensemble method, scholars often apply different models on either the same data or on subsets of data. The outputs from the individual models are then aggregated by averaging or through more complex methodologies.

Hybrid methods integrate the benefits of different methodologies as step-models. For example, the sky image features can be extracted using image-based models. This information is then used as exogenous inputs to train ML models to predict solar power.

Ensemble–Hybrid indicates a combination of ensemble and hybrid methods.

Note that the differences between cooperative ensemble models as defined in [67], hybrid models, and even data processing steps can become vague. We followed the observation's nomenclature, i.e., whenever a paper referred to a model as 'ensemble', 'hybrid', or 'ensemble–hybrid', we grouped it accordingly.

In the database, a categorical variable "ModClass" is created to record the model classification of the forecast model proposed in the reviewed papers.

*Reference model*

As discussed above, the information regarding the reference model used to calculate SS in reviewed papers is included in the database as an independent or explanatory variable. This is a categorical variable named "Reference", including Persistence, SP, and CP.

*Resolution of forecast*

Forecast resolution is represented by the length of each forecasted time step. For example, a forecast of the day-ahead horizon and one-hour resolution is the forecast that predicts the next day with separate values for each hour. Note that the longer the time step is, the lower the resolution, and vice versa. Many scholars believe that a higher resolution such as hourly is more difficult to forecast than a lower resolution such as daily [5]. Therefore, the impact of the resolution on forecast accuracy should also be analyzed. The resolution information measured in minutes is extracted from the reviewed papers and termed "ResMin" in the database.

*Test set length*

The out-of-sample test set (or test set for short) is crucial to validate a forecast model's performance. The variability in the usage of test sets contributes to the complexity of comparing forecasts' accuracy. For example, the test set length has been shown to be negatively correlated with forecast accuracy, possibly due to "cherry-picking" in model assessment or relatively higher uncertainty in a longer period [5]. The impact of test set length on SS is therefore examined in this study. To do this, a numerical variable recording the length of the test set in days is included in the database.

*Train set length*

For many forecast models, training models on a part of the data are required so that models can recognize any pattern of relationship between dependent and explanatory variables. From this training, models can predict the value of the dependent variable based on a new data set of explanatory variables. Therefore, we hypothesize that the amount of training data, usually represented by the length of the train set, can play a role in enhancing forecast accuracy. To examine this, a numerical variable termed "TrainLength" is included in the database to record the length in days of the train set used by reviewed papers.

*Type of forecast*

Solar forecasting includes two key sub-fields, namely, PV output and solar resource forecasting [44]. PV output indicates the electricity generated from solar energy using photovoltaic technology. PV output forecasts, also referred to as PV generation forecasts, are of high interest because they can be directly used by regulators, power plant operators, and industrial stakeholders for PV system planning and operation. In contrast, solar resource forecasts predict solar irradiance, i.e., without translating irradiance into electricity generation. They are important because they have a strong correlation with PV generation [14] and can improve the accuracy of PV forecasting. There is no clear distinction in the methodologies of forecasting either PV or solar resource. Hence, both PV output forecasts and solar resource forecasts are included for the analysis. A categorical variable called "Type" is used to differentiate between the two forecasts in the database.

*Year of Publication*

Many scholars indicate that forecast accuracy improves over time, and a positive correlation between publication date and forecast accuracy is expected. Hence, a numerical variable "Year" is included in the database to track in which year the paper was published.

## Appendix B. Literature identification

Table B.1: Keywords and search sessions.

| No | Date | Keywords | Result number | Note |
|---|---|---|---|---|
| **1** | 27/01/2022 | solar irradiance PV power forecasting "forecast skill" | 1,650 | Stop at p.47 (940 results checked). |
| **2** | 17/02/2022 | solar irradiance PV power forecasting "skill score" | 1,020 | Stop at p.22 (440 results checked). |
| **3** | 28/01/2022 | photovoltaic forecast "skill score" | 1,410 | Stop at p.22 (440 results checked). |
| **4** | 09/02/2022 | photovoltaic "forecast skill" | 515 | All checked. |

A combination of general search and exact search was implemented to identify the literature. The benefit of the general search is to allow Google Scholar to return all the results with synonyms of the keywords. This ensures the results are suitably comprehensive. In searches No. 1–No. 3, the terms for 'solar irradiance', 'PV power', and 'forecasting' were entered as general search terms. Therefore, the results of similar terms (e.g., 'solar radiation', 'PV generation', or 'predicting') were also included in the returns. For all the search sessions, the terms 'skill score' and 'forecast skill' were included as the exact search. This is because the meta-analysis in this paper focuses exclusively on SS, which is also called 'forecast skill' in much of the literature.

For each search session, the search results were preliminarily screened for the relevance based on the titles and the text snippets (the short text description under the title). If the title or the text contains solar resource or PV forecasting equivalent terms, and the term 'skill score' or 'forecast skill' is mentioned, the result is considered relevant. In case of doubt, the result is also included as relevant at this stage. The screening continues until all the results within one result page are not relevant (noise). For search No. 1, for example, all noise was observed starting on page (p.) 47. Therefore, the screening for search No. 1 stopped at p. 47. With Google Scholar, each page covers 20 results, which means the first 940 results of the No. 1 search were screened. The screen then continued in the same manner for the other search sessions. A total of 2,335 search results, with some minor overlapping, were screened. After review, 1,447 unique search results were considered relevant and moved to the next step. The list of 1,447 literatures is provided in the supplementary data file.

## Appendix C. 188 papers for data extraction

Table C.1: 70 papers on PV output forecasting. The table is sorted alphabetically based on the last name of the first author.

| No | Ref | No | Ref |
|---|---|---|---|
| **1** | [68] | **36** | [69] |
| **2** | [70] | **37** | [71] |
| **3** | [72] | **38** | [73] |
| **4** | [74] | **39** | [75] |
| **5** | [76] | **40** | [77] |
| **6** | [78] | **41** | [79] |
| **7** | [80] | **42** | [81] |
| **8** | [82] | **43** | [83] |
| **9** | [84] | **44** | [85] |
| **10** | [86] | **45** | [87] |
| **11** | [88] | **46** | [89] |
| **12** | [90] | **47** | [91] |
| **13** | [92] | **48** | [93] |
| **14** | [94] | **49** | [95] |
| **15** | [96] | **50** | [97] |
| **16** | [98] | **51** | [99] |
| **17** | [100] | **52** | [101] |
| **18** | [102] | **53** | [103] |
| **19** | [104] | **54** | [105] |
| **20** | [106] | **55** | [107] |
| **21** | [108] | **56** | [109] |
| **22** | [110] | **57** | [111] |
| **23** | [112] | **58** | [113] |
| **24** | [114] | **59** | [115] |
| **25** | [116] | **60** | [117] |
| **26** | [118] | **61** | [119] |
| **27** | [120] | **62** | [121] |
| **28** | [122] | **63** | [123] |
| **29** | [124] | **64** | [125] |
| **30** | [126] | **65** | [127] |
| **31** | [128] | **66** | [129] |
| **32** | [130] | **67** | [131] |
| **33** | [132] | **68** | [133] |
| **34** | [134] | **69** | [135] |
| **35** | [136] | **70** | [137] |

Table C.2: 118 papers on solar resource forecasting.

| No | Ref | No | Ref |
|---|---|---|---|
| **1** | [138] | **60** | [139] |
| **2** | [140] | **61** | [141] |
| **3** | [142] | **62** | [143] |
| **4** | [144] | **63** | [145] |
| **5** | [146] | **64** | [147] |
| **6** | [148] | **65** | [149] |
| **7** | [150] | **66** | [151] |
| **8** | [152] | **67** | [153] |
| **9** | [154] | **68** | [155] |
| **10** | [156] | **69** | [157] |
| **11** | [158] | **70** | [159] |
| **12** | [160] | **71** | [161] |
| **13** | [162] | **72** | [163] |
| **14** | [164] | **73** | [165] |
| **15** | [166] | **74** | [167] |
| **16** | [168] | **75** | [169] |
| **17** | [170] | **76** | [171] |
| **18** | [172] | **77** | [173] |
| **19** | [174] | **78** | [175] |
| **20** | [176] | **79** | [177] |
| **21** | [178] | **80** | [179] |
| **22** | [180] | **81** | [181] |
| **23** | [182] | **82** | [183] |
| **24** | [184] | **83** | [185] |
| **25** | [186] | **84** | [187] |
| **26** | [188] | **85** | [189] |
| **27** | [190] | **86** | [191] |
| **28** | [23] | **87** | [192] |
| **29** | [193] | **88** | [194] |
| **30** | [195] | **89** | [196] |
| **31** | [197] | **90** | [198] |
| **32** | [199] | **91** | [200] |
| **33** | [201] | **92** | [202] |
| **34** | [203] | **93** | [204] |
| **35** | [205] | **94** | [206] |
| **36** | [207] | **95** | [208] |
| **37** | [209] | **96** | [210] |
| **38** | [211] | **97** | [212] |
| **39** | [213] | **98** | [214] |
| **40** | [215] | **99** | [216] |
| **41** | [217] | **100** | [218] |
| **42** | [219] | **101** | [220] |
| **43** | [221] | **102** | [222] |
| **44** | [223] | **103** | [48] |
| **45** | [224] | **104** | [225] |
| **46** | [226] | **105** | [227] |
| **47** | [228] | **106** | [229] |
| **48** | [230] | **107** | [231] |
| **49** | [232] | **108** | [233] |
| **50** | [234] | **109** | [235] |
| **51** | [236] | **110** | [237] |
| **52** | [238] | **111** | [239] |
| **53** | [240] | **112** | [241] |
| **54** | [242] | **113** | [243] |
| **55** | [244] | **114** | [245] |
| **56** | [246] | **115** | [247] |
| **57** | [62] | **116** | [248] |
| **58** | [249] | **117** | [250] |
| **59** | [251] | **118** | [252] |

## Appendix D. Database overview

Table D.1: Statistics summary of numerical and dummy variables. N=Number of observations; SD=Standard deviation; Trim=Trimmed mean value; MAD=Median absolute deviation; SE=Standard error

| **Variable** | **N** | **Mean** | **SD** | **Median** | **Trim** | **MAD** | **Min** | **Max** | **Range** | **Skew** | **Kurtosis** | **SE** |
|---|---|---|---|---|---|---|---|---|---|---|---|---|
| Horizon | 4687 | 497.95 | 823.80 | 120 | 302.29 | 88.96 | 0.17 | 4320.00 | 4319.83 | 2.02 | 3.10 | 12.03 |
| InputHist* | 4687 | 0.91 | 0.29 | 1 | 1.00 | 0.00 | 0.00 | 1.00 | 1.00 | -2.81 | 5.92 | 0.00 |
| InputMete* | 4687 | 0.53 | 0.50 | 1 | 0.53 | 0.00 | 0.00 | 1.00 | 1.00 | -0.11 | -1.99 | 0.01 |
| InputNWP* | 4687 | 0.19 | 0.39 | 0.00 | 0.11 | 0.00 | 0.00 | 1.00 | 1.00 | 1.59 | 0.53 | 0.01 |
| InputST* | 4687 | 0.23 | 0.42 | 0 | 0.17 | 0.00 | 0.00 | 1.00 | 1.00 | 1.27 | -0.40 | 0.01 |
| ResMin | 4687 | 52.39 | 50.63 | 60 | 49.32 | 0.00 | 0.02 | 360.00 | 359.98 | 4.68 | 26.51 | 0.74 |
| SS | 4687 | 23.90 | 21.09 | 21.45 | 23.30 | 17.12 | -94.61 | 96.10 | 190.71 | -0.41 | 3.89 | 0.31 |
| TestLength | 4687 | 303.51 | 195.55 | 350 | 279.77 | 113.05 | 1.00 | 1460.00 | 1459.00 | 1.35 | 3.98 | 2.86 |
| TrainLength | 4687 | 798.22 | 1114.54 | 396 | 560.43 | 450.71 | 0.00 | 7305.00 | 7305.00 | 3.64 | 15.71 | 16.28 |
| Year | 4687 | 2018.19 | 2.25 | 2018 | 2018.20 | 2.97 | 2006.00 | 2022.00 | 16.00 | -0.19 | -0.65 | 0.03 |
| *Note: *: dummy variable, otherwise: numerical variable* | | | | | | | | | | | | |

Table D.1 presents a statistics summary for numerical and dummy variables. The statistics summary is important to get an overview of a variable. For example, the statistics for the SS variable show that the average value (Mean) of SS in the database is 23.90% and the median is 21.09%. The average SS after removing the outliers (Trim) is 23.30%, which is not far off the normal average. This indicates a low impact of outliers in the database. The minimum SS is -94.61 and the maximum SS is 96.10, resulting a range of 190.71. The distribution is left-skewed (negative Skew value) with heavy tails (positive Kurtosis). Other information on the variable, such as standard error (SE), standard deviation (SD), and median absolute deviation (MAD), is also presented in the table.

For dummy variables, the mean value also shows the proportion of data where the dummy equals 1. For instance, the mean values of InputHist is 0.91. This means 91% of the data use historical power data for the inputs.

## Appendix E. MARS cross-validation and diagnostic plots

A ten-fold cross-validation grid search was conducted to identify the optimal combination of two hyperparameters that are associated with MARS: the highest degree of interactions and the maximum number of terms retained in the final model. For degree of interactions, we limit the value to 2 degrees so that the interpretation of the results is possible. Regarding the maximum number of terms, values in a range from 2 to 100, with a step of 10, were included in the grid search. The values of the hyperparameters that minimizes the prediction error (RMSE) of the model were selected. The best value for degree of interactions was 2, and for number of terms was 34, as can be seen from Figure E.1.

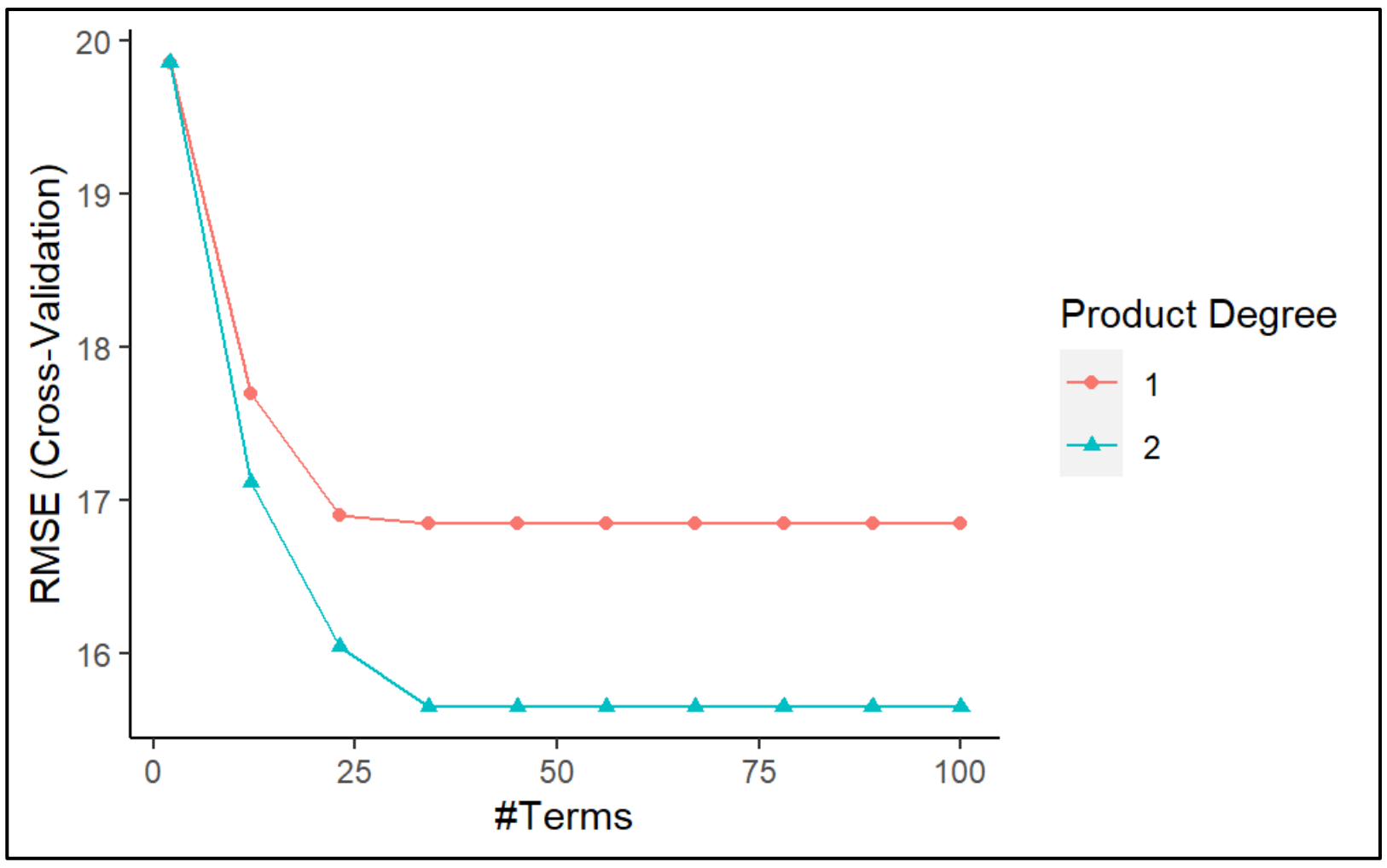

Figure E.1: Ten-fold cross-validation for MARS

The diagnostic plots of the regression is provided in Figure E.2. Plot a presents the process used to identify the optimal number of terms (hinge functions) and variables or predictors. For each iteration, RSq ($R^2$) and GRSq (Generalized $R^2$) are recorded. The model with the highest GRSq is selected. The plot shows that 28 terms and 14 predictors are included in the best model. The $R^2$ is 45.98% and the generalized $R^2$ is 44.39%.

Plots b through d analyze the residuals of the model. Residuals are estimates of models' errors. Therefore, any patterns in the data left unexplained by the model are observed the residuals. For a good model, the residuals are expected to have normal and independent distribution. Plots b and c show that this property of the residuals is well-satisfied in the model. In plot b, the absolute value of the model residuals [abs(Residuals)] starts at zero and quickly rises to one, indicating a normal distribution. Plot c shows that no pattern is observed in the residuals. In plot d, the distribution of the residual is compared to a normal distribution. The largest deviations are the outliers.

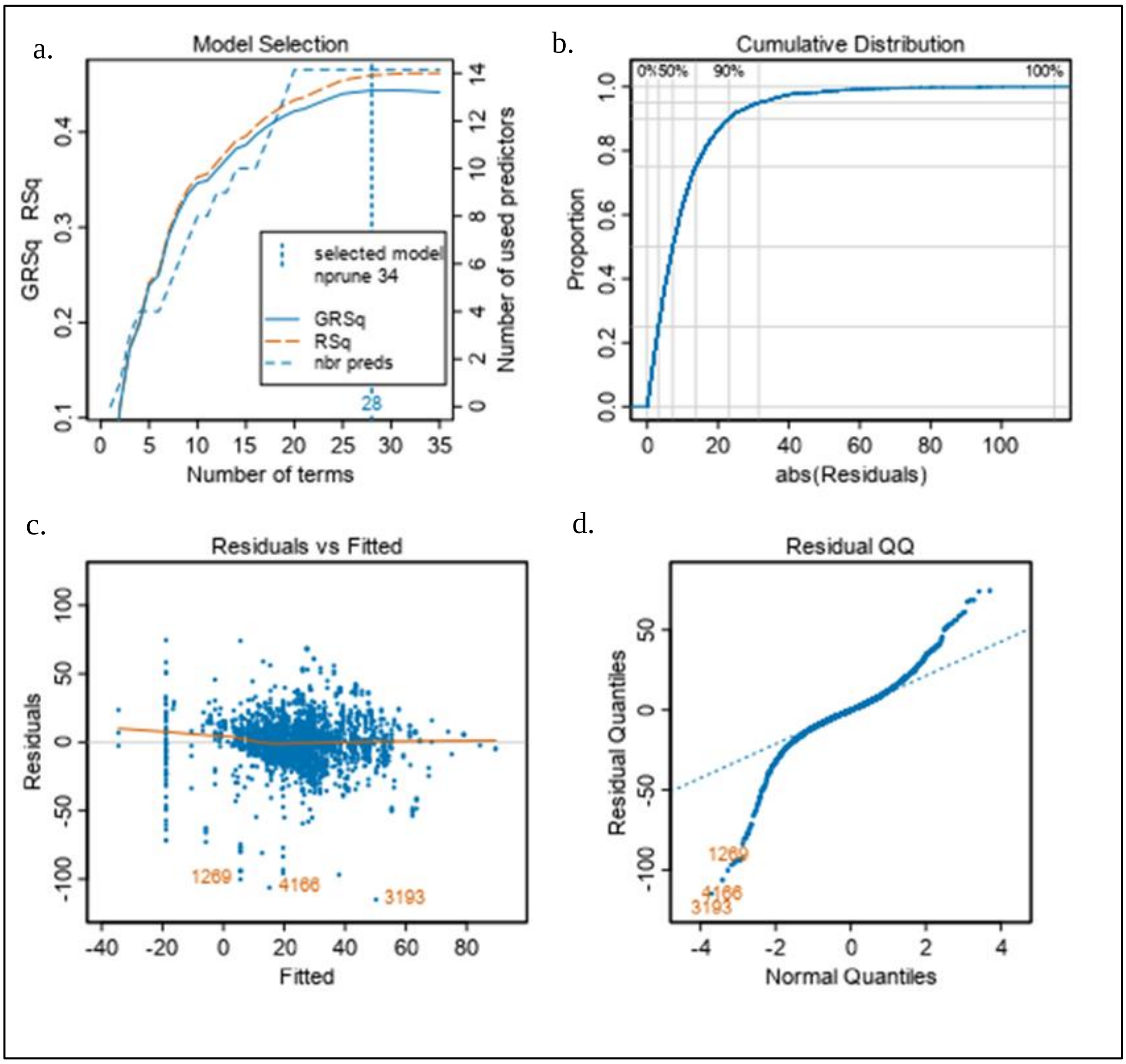

Figure E.2: Diagnostic plots for MARS.